%% file: main.tex
\documentclass[ALICE,manyauthors]{cernphprep}

\usepackage[comma,square,numbers,sort&compress]{natbib}
\usepackage{hyperref}
\usepackage{lineno}
\usepackage{xcolor}
\usepackage{textcomp}
\usepackage[T1]{fontenc}
\usepackage{amsmath}
\usepackage{slashed}
\usepackage{setspace}
\usepackage{orcidlink}
\input{commands}

\begin{document}%

\begin{titlepage}
\PHyear{2022}
\PHnumber{188}      
\PHdate{15 September}  
%

\title{Measurement of the lifetime and $\Lambda$ separation energy of \hyp}
\ShortTitle{Measurement of the lifetime and $\Lambda$ separation energy of \hyp}   
\hyphenation{nu-cleo-syn-the-sis}
\Collaboration{ALICE Collaboration\thanks{See Appendix~\ref{app:collab} for the list of collaboration members}}
\ShortAuthor{ALICE Collaboration} 

\begin{abstract}
\input{abstract}
\end{abstract}
\end{titlepage}
\setcounter{page}{2}

\input{text}

\newenvironment{acknowledgement}{\relax}{\relax}
\begin{acknowledgement}
\section*{Acknowledgements}
\input{fa_2022-08-31_Opt_C.tex}    
\end{acknowledgement}

\bibliographystyle{utphys}   
\bibliography{biblio}


\appendix
\newpage
\input{appendix}               
\newpage

\section{The ALICE Collaboration}
\label{app:collab}
 \input{2022-08-31-Alice_Authorlist_2022-08-31_Opt_C.tex}  
\end{document}

%% file: commands.tex
\usepackage{xspace}

\newcommand{\PbPb}         {\mbox{Pb--Pb}\xspace}

\newcommand{\snn}          {\ensuremath{\sqrt{s_{\mathrm{NN}}}}\xspace}
\newcommand{\pt}           {\ensuremath{p_{\rm T}}\xspace}
\newcommand{\ct}           {\ensuremath{ct}\xspace}
\newcommand{\BL}           {\ensuremath{\mathrm{B_{\Lambda}}}\xspace}
\newcommand{\muhyp}         {\ensuremath{\mu_{^{3}_{\Lambda}\mathrm H}}\xspace}

\newcommand{\dEdx}         {\ensuremath{\textrm{d}E/\textrm{d}x}\xspace}

\newcommand{\nineH}        {$\sqrt{s}~=~0.9$~Te\kern-.1emV\xspace}
\newcommand{\seven}        {$\sqrt{s}~=~7$~Te\kern-.1emV\xspace}
\newcommand{\twoH}         {$\sqrt{s}~=~0.2$~Te\kern-.1emV\xspace}
\newcommand{\twosevensix}  {$\sqrt{s}~=~2.76$~Te\kern-.1emV\xspace}
\newcommand{\five}         {$\sqrt{s}~=~5.02$~Te\kern-.1emV\xspace}
\newcommand{\twosevensixnn}{$\sqrt{s_{\mathrm{NN}}}~=~2.76$~Te\kern-.1emV\xspace}
\newcommand{\fivenn}       {$\sqrt{s_{\mathrm{NN}}}~=~5.02$~Te\kern-.1emV\xspace}

\newcommand{\GeVc}         {Ge\kern-.1emV/$c$\xspace}
\newcommand{\MeVc}         {Me\kern-.1emV/$c$\xspace}
\newcommand{\TeV}          {Te\kern-.1emV\xspace}
\newcommand{\GeV}          {Ge\kern-.1emV\xspace}
\newcommand{\MeV}          {Me\kern-.1emV\xspace}
\newcommand{\GeVmass}      {Ge\kern-.2emV/$c^2$\xspace}
\newcommand{\MeVmass}      {Me\kern-.2emV/$c^2$\xspace}

\newcommand{\Hee}          {\ensuremath{^{3}\mathrm{He}}\xspace}
\newcommand{\antiHee}      {\ensuremath{^{3}\overline{\mathrm{He}}}\xspace}
\newcommand{\hyp}          {\ensuremath{^{3}_{\Lambda}\mathrm H}\xspace}
\newcommand{\antihyp}      {\ensuremath{^{3}_{\overline{\Lambda}} \overline{\mathrm H}}\xspace}

%% file: abstract.tex
The most precise measurements to date of the \hyp lifetime $\tau$ and $\Lambda$ separation energy \BL are obtained using the data sample of Pb--Pb collisions at $\snn=$ 5.02 TeV collected by ALICE at the LHC. The \hyp is reconstructed via its charged two-body mesonic decay channel (\hyp $\rightarrow$ $^3$He + $\pi^-$ and the charge-conjugate process). The measured values $\tau = [253 \pm 11 \text{~(stat.)} \pm 6 \text{~(syst.)}]$ ps and $\BL= [102 \pm 63 \text{~(stat.)} \pm 67 \text{~(syst.)}]$ keV are compatible with predictions from effective field theories and confirm that the \hyp structure is consistent with a weakly-bound system.

%% file: text.tex
Hypernuclei are bound states of nucleons and hyperons that are particularly interesting because they can be used as experimental probes for the study of the hyperon--nucleon (Y--N) interaction.
Searching for hypernuclei and exploring the Y--N interaction have been a source of fascination for nuclear physicists since the discovery of the first hypernuclei in 1953~\cite{Danysz}. In recent years,  measurements of the hypertriton production and lifetime have stimulated an interesting debate in the high-energy physics community. The knowledge of the Y--N interaction has become more relevant recently due to its connection to the modelling of dense astrophysical objects like neutron stars~\cite{schaffner-bielich_2020, Lattimer:2004pg}. Indeed, in the inner core of neutron stars the creation of hyperons is energetically favoured compared to purely nucleonic matter~\cite{Tolos:2020aln}. The presence of hyperons as additional degrees of freedom leads to a considerable modification of the matter equation of state (EOS), prohibiting the formation of high-mass neutron stars. This is incompatible with the observation of neutron stars heavier than two solar masses~\cite{schaffner-bielich_2020, Tolos:2020aln}, constituting what is referred to as the "hyperon puzzle". Many attempts were made to solve this puzzle, e.g. by introducing three-body forces leading to an additional repulsion that can counterbalance the large gravitational pressure and allow for larger star masses~\cite{Lonardoni:2014bwa,Logoteta:2019utx}. To constrain the parameter space of such models, a detailed knowledge of the Y--N interaction and of the three-body Y--N--N interaction is mandatory, including $\Lambda$, $\Sigma$, and $\Xi$ hyperons. Numerous particle correlation analyses~\cite{Fabbietti:2020bfg,ALICE:2020mfd} directly contribute to the determination of such interactions. In a complementary approach, the lifetime and the binding energy of a hypernucleus reflect the strength of the Y--N interaction~\cite{Dalitz:1962eb, Kamada:1997rv}.
The current estimate of the separation energy of the $\Lambda$ in the hypertriton is $\BL =$ $181\pm48$ keV~\cite{HypernuclearDataBase}, which results in a RMS radius (average distance of the $\Lambda$ to the deuteron) of the order of 10 fm~\cite{riisager2000quantum, Braun-Munzinger:2018hat, Gal:2018bvq, Hildenbrand:2019sgp}. 
Lower values ($\sim 90$ keV) of \BL are favoured when fitting the correlation functions for protons and $\Lambda$ baryons~\cite{ALICE:2018ysd, Haidenbauer:2019boi, ALICE:2021njx}, therefore new measurements are required to understand this tension.
The latest theoretical calculations predict a different degree of dependence of the \hyp lifetime on its binding energy. For pionless effective field theory (EFT)~\cite{Hildenbrand:2020kzu} based calculations, the \hyp lifetime is very close to the free $\Lambda$ lifetime, with very little binding energy dependence for \BL values spanning from 0 to 0.5 MeV. On the other hand, in the $\chi$EFT approach~\cite{Perez-Obiol:2020qjy}, a stronger dependence on the binding energy is predicted for the \hyp lifetime being $\tau=(163\pm18)$ ps for $\BL=410$ keV and $\tau=(234\pm27)$ ps for $\BL = 69$ keV.
Previous measurements of the lifetime~\cite{STAR:2010gyg,Rappold:2013fic,Adam:2015yta,Adamczyk:2017buv,Acharya:2019qcp,STAR:2021orx} and \BL~\cite{Adam:2019phl} of \hyp in heavy-ion collisions have still quite large uncertainties.
In this Letter, new measurements with unprecedented precision of the \hyp lifetime and binding energy are presented to address the questions about its structure. 

The presented results are based on data collected during the  2018 \PbPb LHC run at a centre-of-mass
energy per nucleon pair of \snn $=$ 5.02 TeV. The ALICE detector and its performance are described in detail in Refs.~\cite{Aamodt:2008zz,Abelev:2014ffa}.
The data acquisition for \PbPb events is triggered by the V0A and V0C scintillation detectors~\cite{Abbas:2013taa}, positioned at forward ($2.8<\eta<5.1$) and backward ($-3.7 <\eta <-1.7$) pseudorapidity, respectively.
A coincidence of signals in both V0A and V0C is used as a minimum-bias trigger. In addition, two thresholds on the minimum amount of charge deposited on the V0 detector are employed to trigger on central and semi-central \PbPb collisions.
A centrality estimator based on the V0 detector arrays~\cite{ALICE-PUBLIC-2018-011} is used to  select the 90\% most central hadronic collisions. 
Events are further selected by allowing a 10 cm maximum displacement of the primary vertex along the beam axis from the nominal centre of the experiment in order to benefit from the full acceptance of the detector. Finally, events with multiple reconstructed primary vertices are rejected to avoid ambiguous associations of \hyp candidates to their production vertices.
In total, about 300 million events are selected for this analysis. 

In \PbPb collisions at the LHC, approximately the same number of \hyp and \antihyp are expected to be produced. The \hyp decays are detected via the charged two-body channel $\hyp \rightarrow \mathrm{^3He} + \pi^-$ (and the corresponding charge-conjugated particles for \antihyp). The decay products of the \hyp are tracked  with the Inner Tracking System~\cite{Aamodt:2010aa} and the Time Projection Chamber (TPC)~\cite{Alme:2010ke}, which are positioned within a solenoid providing a homogeneous magnetic field of $0.5$ T in the direction of the beam axis. Charged particles are tracked over the full azimuth and in the pseudorapidity interval $|\eta| < 0.8$.
The specific energy loss of the decay products of \hyp and \antihyp is also measured in the TPC, with a \dEdx resolution of about $5\%$~\cite{Alme:2010ke, ALICE:2014juv}.
The $n(\sigma^{\rm TPC}_{i})$ variable represents the particle identification (PID) response in the TPC expressed in terms of the deviation between the measured and the expected \dEdx for a particle species $i$, in units of the detector resolution $\sigma$. The expected \dEdx is computed with a parametrised Bethe--Bloch function~\cite{Abelev:2014ffa}. Pion and \Hee tracks within $\pm 5\sigma^{\rm TPC}$ are selected.

The identified \Hee and $\pi$ tracks are then used to reconstruct the \hyp weak-decay topology with an algorithm similar to the one employed in previous analyses~\cite{Adam:2015yta, Acharya:2019qcp, ALICE:2021sdc}. By combining the information on the decay kinematics and decay vertex, several selection variables are defined. Those used in the analysis are the following: the distance of closest approach  both from the primary and the decay vertex and the $n(\sigma^{\rm TPC}_{i})$ of each daughter track, the number of clusters of the \Hee track in the TPC, the reconstructed \pt of the \hyp, and cos($\mathrm{\theta_P}$), where $\mathrm{\theta_P}$ is the angle between the total momentum vector of the decay daughters and the straight line connecting the primary and secondary vertices.
These variables are combined as a gradient-boosted decision tree classifier (BDT)~\cite{10.1145/2939672.2939785, hipe4ml} that is trained on a dedicated Monte Carlo (MC) simulated event sample.
The MC sample consists of \hyp (decays) signals injected onto underlying
\PbPb collisions simulated with the \textsc{HIJING} event generator~\cite{Wang:1991hta}. The
transverse momentum (\pt) distribution of the injected signals is given
by the blast-wave~\cite{Schnedermann:1993ws} function, with parameters taken from simultaneous fits of the \pt distribution of light-flavoured hadrons measured in \PbPb collisions~\cite{ALICE:2019hno}.
Only candidates with $2\leq \pt < 9$ GeV/$c$ are considered.
The particle transport through the detector material is done using \textsc{Geant4}~\cite{Agostinelli:2002hh}, which simulates the interaction with the material and the weak decay of the \hyp.
The BDT is a supervised learning algorithm that determines how to discriminate between two or more classes, in this case signal and background, by
examining sets of examples called the training sets. In this analysis, the training sets comprise \hyp signal candidates extracted from the MC sample and background candidates from paired like-sign $\mathrm{^3He}$ and $\mathrm{\pi}$ tracks from data.
For each \hyp candidate, the BDT combines topological and single-track variables to return a score, which is used to discriminate between signal or background. In particular, the most important variable employed by the BDT for the classification is the cos($\mathrm{\theta_P}$). Candidates with a BDT score higher than a given threshold are
selected as signal. The threshold is defined to maximise the expected signal significance assuming a production yield as predicted by the thermal statistical hadronization model~\cite{Braun-Munzinger:2018hat} for the \hyp and the background rate observed when combining like-sign \Hee and $\pi$ pairs.

\begin{figure*}[t]
    \centering
    \includegraphics[width=0.49\textwidth]{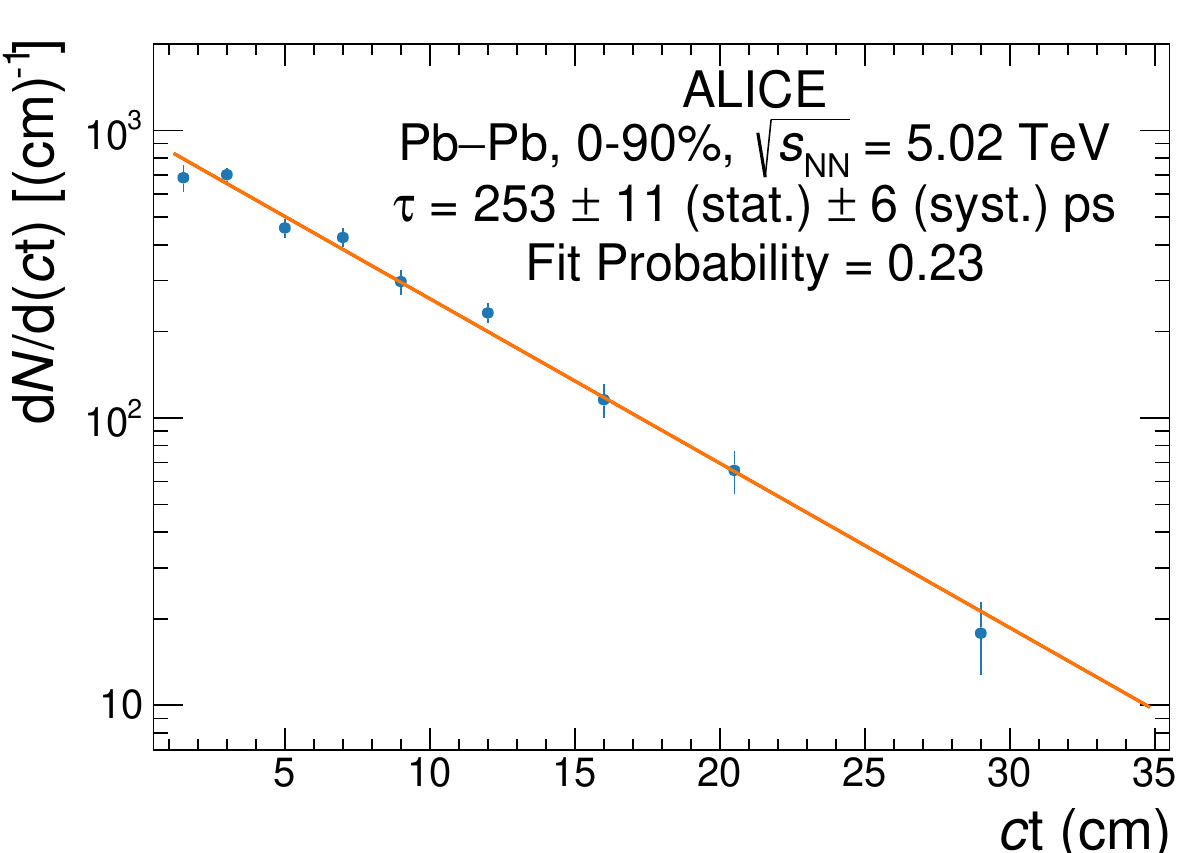}
    \includegraphics[width=0.49\textwidth]{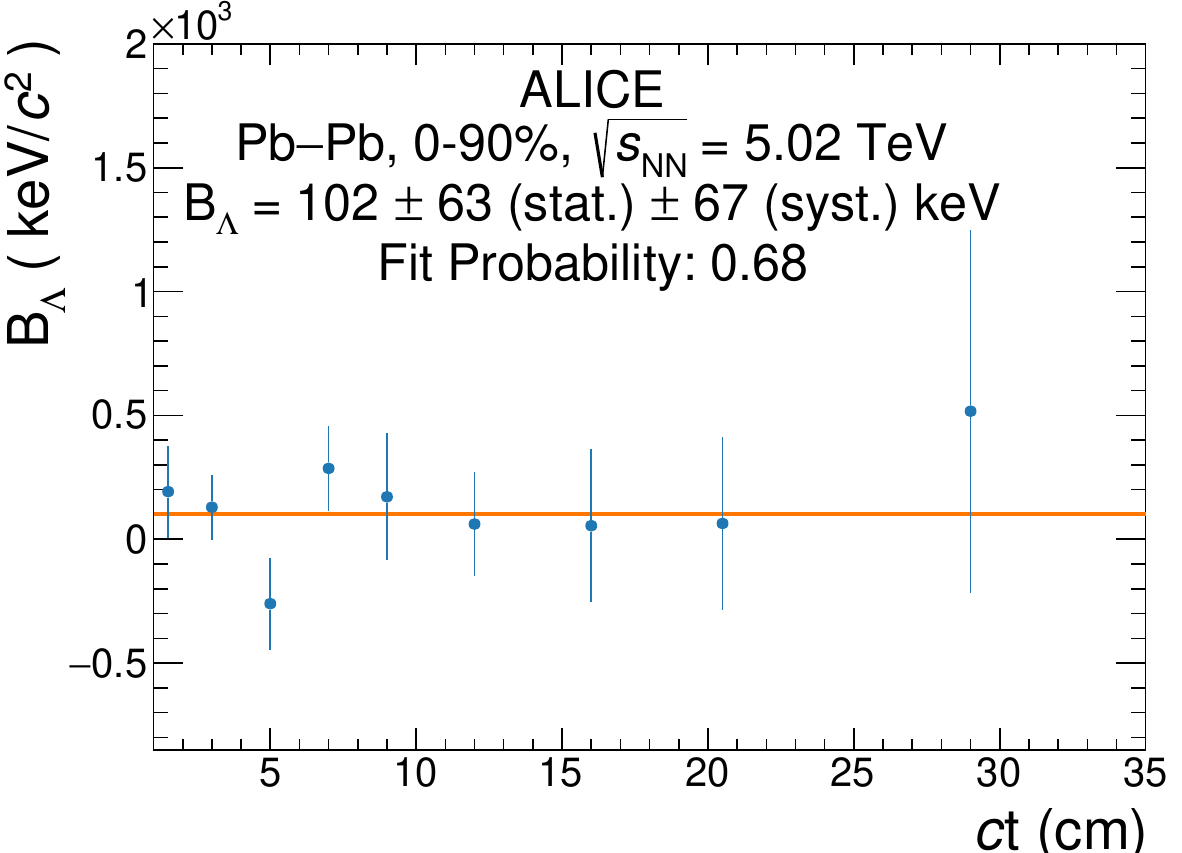}
    \caption{Left: exponential decay spectrum as a function of the proper decay length for \hyp, the blue points represent the measured yield, while the orange line represents the best fit to the measurement. Right: \BL measurement as a function of the proper decay length. Only statistical uncertainties are shown; see the text for a description of the determination of the systematic uncertainties. The fit probability computed with a Pearson test is reported.}
    \label{fig:final}
\end{figure*}

The \hyp lifetime is extracted by analysing the proper decay length spectrum shown in Fig.~\ref{fig:final}. The sample of \antihyp and \hyp candidates is divided into nine $\ct = ML/p$ intervals, where $c$ is the speed of light, $t$ is the proper time of the candidate, $M$ is the mass
of the candidate, $L$ is the decay distance, and $p$ is the reconstructed
momentum. The BDT training and threshold optimisation is repeated for each \ct interval.
The candidates that pass the BDT selection are used to populate the invariant-mass ($m=\sqrt{(E_\pi+E_{\Hee})^2-|\vec{p}_{\pi}+\vec{p}_{\Hee}|^2}$) distributions, as shown in Appendix~\ref{suppmat}. An unbinned maximum-likelihood fit is performed on the invariant-mass distribution using a Kernel Density Estimator (KDE) function~\cite{Cranmer:2000du, Verkerke:2003ir}, constructed using the MC sample to describe the signal and a linear function to describe the background. The distributions obtained in
the nine \ct intervals can be found in Appendix~\ref{suppmat}. The KDE is used to model the non-Gaussian behaviour of the signal shape observed in the simulation by means of a superposition of Gaussian functions. The widths of the Gaussian functions are determined with an adaptive procedure that takes into account the local density of the signal in the MC sample. 
The yield in each \ct interval is obtained from the fit to the invariant-mass spectrum. The fitted signal is corrected for the reconstruction and selection efficiency, 
and for the acceptance of the ALICE detector. While the interactions of the daughter particles are correctly reproduced by \textsc{Geant4}~\cite{ALICE:2022zuz} and therefore naturally accounted in the efficiency determination, the interaction of the \hyp and \antihyp requires a dedicated treatment. According to Ref.~\cite{Evlanov:1998py}, the expected absorption cross section of \hyp due to the inelastic interactions in the ALICE detector material is about 1.5 times that of \antiHee ($\mathrm{\sigma_{inel}^{\antiHee}}$). This value is used for simulating the passage of \hyp in the detector and evaluating the effect on the reconstruction efficiency. Different corrections have been applied for \hyp and \antihyp according to their expected cross sections.
The typical total efficiency (including acceptance, reconstruction, and selection) is around 15\% while the absorption probability of \hyp due to the inelastic interactions in the ALICE detector material is of the order of a few percent. The obtained spectrum is fitted with an exponential function as shown in the left panel of Fig.~\ref{fig:final}. The fit is performed by using the integral of the function in each bin to account for the variable widths of the \ct intervals.

The major systematic uncertainties come from (1) the \hyp identification and, (2) the uncertainty on the \hyp inelastic interactions in the detector. The total uncertainty is obtained as the quadratic sum of the individual contributions. The first, dominant contribution is computed by varying simultaneously the BDT selection efficiency ($\pm 10\%$), the background shape (linear, second order polynomial, and exponential) and the signal (KDE and double sided Crystal Ball~\cite{ALICE-PUBLIC-2015-006}) fit functions in each \ct interval. The systematic uncertainty is given by the RMS of the distribution of lifetimes obtained from $5\times10^{4}$ different combinations and amounts to approximately 2\%.
The second contribution is evaluated by varying the \hyp absorption cross section and evaluating the effect on the lifetime. 
The systematic uncertainty due to the assumption on the \hyp absorption cross section is evaluated by employing different cross sections for the \hyp from zero (no interaction) to $2\mathrm{\sigma_{inel}^{\antiHee}}$. For each variation the lifetime is recalculated, resulting in a systematic uncertainty of 1\%.

The $\Lambda$ separation energy \BL is obtained in each \ct interval using the \hyp mass (\muhyp) extracted from the fit (see Appendix~\ref{suppmat}), the deuteron mass taken from CODATA~\cite{CODATA}, and the $\Lambda$ mass taken from the PDG~\cite{Zyla:2020zbs}. 
The reconstructed value of \muhyp is affected by the imperfect correction for the energy loss of the daughter particles in the ALICE material. This effect produces a shift that depends on the radial distance travelled by the \hyp candidates before decaying, and it is evaluated by analysing the MC simulations ($\delta_{\mathrm{MC}}$). 
The values of $\delta_{\mathrm{MC}}$ as a function of the $\ct$ of the decay particles are shown in Appendix~\ref{suppmat},
and they span between -0.1 MeV and 0.8 MeV.
To account for the possible mismatch between data and simulation, an additional data-driven correction is applied based on a dedicated precise measurement of the $\Lambda$ mass. This represents an ideal test for the full analysis chain and for a potential data--simulation mismatch since the $\Lambda$ mass is known with a precision of 6 keV~\cite{Zyla:2020zbs} and its lifetime is compatible within $1\sigma$ with the \hyp one. Hence, the $\Lambda$ mass is computed with the same analysis procedure employed for the \hyp, and the positive shift obtained with respect to the PDG value ($\delta_{\Lambda} = 66$ keV) is used as our estimate of data--simulation mismatch and for correcting our value of \muhyp. Finally, in each \ct interval \BL is computed as $\BL = m_{\mathrm{d}} + m_{\Lambda} - (\muhyp - \delta_{\mathrm{MC}} -  \delta_{\Lambda})$. 
The final \BL value and its statistical uncertainty are obtained from the average of the values measured in each $ct$ interval (see Fig.~\ref{fig:final}, right) weighted on their statistical uncertainties.

The systematic uncertainties on \BL originate from (1) the \hyp selection and the signal extraction, (2) the uncertainty on $\delta_{\Lambda}$, and (3) the uncertainty on the ALICE material budget. The first contribution is computed using the same method as for the lifetime analysis and amounts to $\pm$ 29 keV. The uncertainty on $\delta_{\Lambda}$ takes into account a 60 keV shift observed by repeating the mass measurement for $\Lambda$ and $\overline{\Lambda}$ with different magnetic field polarities. Finally, the systematic contribution due to the uncertainty of the ALICE material budget is computed by varying the material budget in the MC sample by its uncertainty and repeating the analysis. The \BL is recomputed for each variation, resulting in a systematic uncertainty of 8 keV.
The systematic uncertainty is taken as the sum in quadrature of the three contributions. For the determination of the mean lifetime of the \hyp and its \BL, the contribution from the knowledge of the magnetic field is considered by performing the analysis separately for positive and negative polarities of the solenoidal magnet. As the analyses with the two polarities returned results statistically compatible with each other, no further systematic uncertainty is added.

For both the lifetime and the \BL analyses, other potential sources of systematic uncertainty were tested, such as the input \pt and \ct shape of \hyp in the Monte Carlo sample, the BDT hyperparameters, the discrepancy between BDT and linear selections, and the \hyp reconstruction algorithm, all resulting in a non-significant contribution.

\begin{figure}[t]
    \centering
    \includegraphics[width=0.5\textwidth]{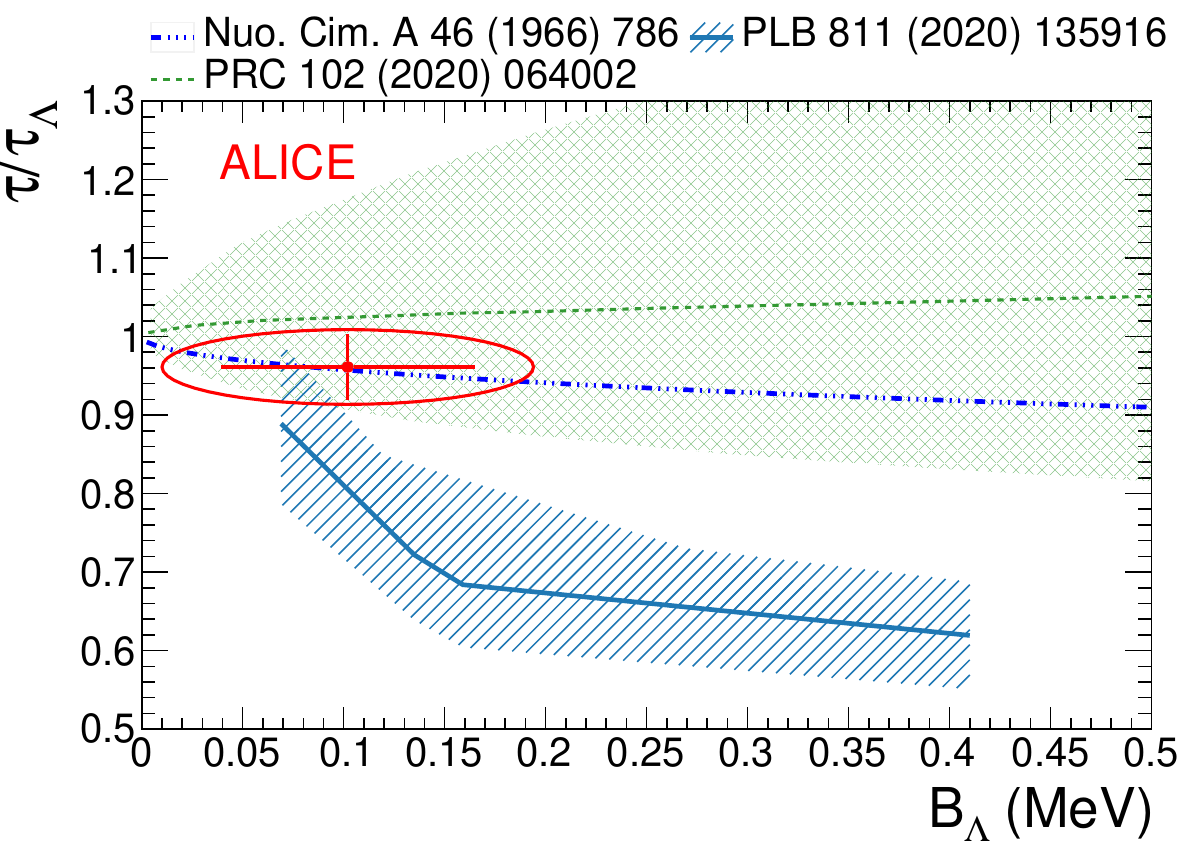}
    \caption{The \hyp lifetime relative to the free $\Lambda$ lifetime as a function of the \BL for pionless EFT~\cite{Hildenbrand:2020kzu} (green),
    $\chi$EFT~\cite{Perez-Obiol:2020qjy} (light blue), and the original $\pi$ exchange calculations~\cite{Rayet:1966fe} (blue). The red point represents the measurement presented in this Letter with the statistical and total uncertainties depicted with lines and an ellipse, respectively.}
    \label{fig:2dmeas}
\end{figure}
The measurements for the \hyp and \antihyp lifetime and \BL obtained with this analysis are
\[
\tau = [253 \pm 11 \text{~(stat.)} \pm 6 \text{~(syst.)]~ps},
\]
\[
\BL= [102 \pm 63 \text{~(stat.)} \pm 67 \text{~(syst.)]~keV.}
\]

As shown in Fig.~\ref{fig:2dmeas}, the measurements are in agreement with both the predictions from pionless EFT~\cite{Hildenbrand:2020kzu} and $\chi$EFT~\cite{Perez-Obiol:2020qjy}, while they severely restrict the phase space available for these theories and strongly confirm the weakly-bound nature of \hyp.
Furthermore, the new measurement of the \BL is in agreement within 1$\sigma$ with the binding energy value describing best the p--$\Lambda$ correlations measured with the femtoscopy technique~\cite{Haidenbauer:2019boi, ALICE:2021njx}.

Finally, the relative differences between the \hyp and \antihyp lifetimes and masses are measured, giving the values

\[
\frac{\tau_{\hyp}-\tau_{\antihyp}}{\tau_{\hyp}} = [3 \pm 7\text{~(stat.)} \pm 4\text{~(syst.)]}\times 10^{-2},
\]
\[
    \frac{m_{\hyp}-m_{\antihyp}}{m_{\hyp}} = [5 \pm 5\text{~(stat.)} \pm 3
    \text{~(syst.)}] \times 10^{-5},
\]

which are consistent with zero and, therefore, with the CPT symmetry expectation. Note, in the mass difference measurement, the decay daughter masses
are taken to be the same between particles and antiparticles.

In summary, the most precise measurements to date of the \hyp lifetime and \BL, presented in this Letter, strongly support the loosely-bound nature of \hyp. The measured value perfectly agrees with the \BL that best fits the correlation functions for protons and $\Lambda$ baryons within the current theoretical approaches~\cite{ALICE:2018ysd, Haidenbauer:2019boi, ALICE:2021njx}. 

\begin{figure*}[t]
    \centering
    \includegraphics[width=0.49\textwidth]{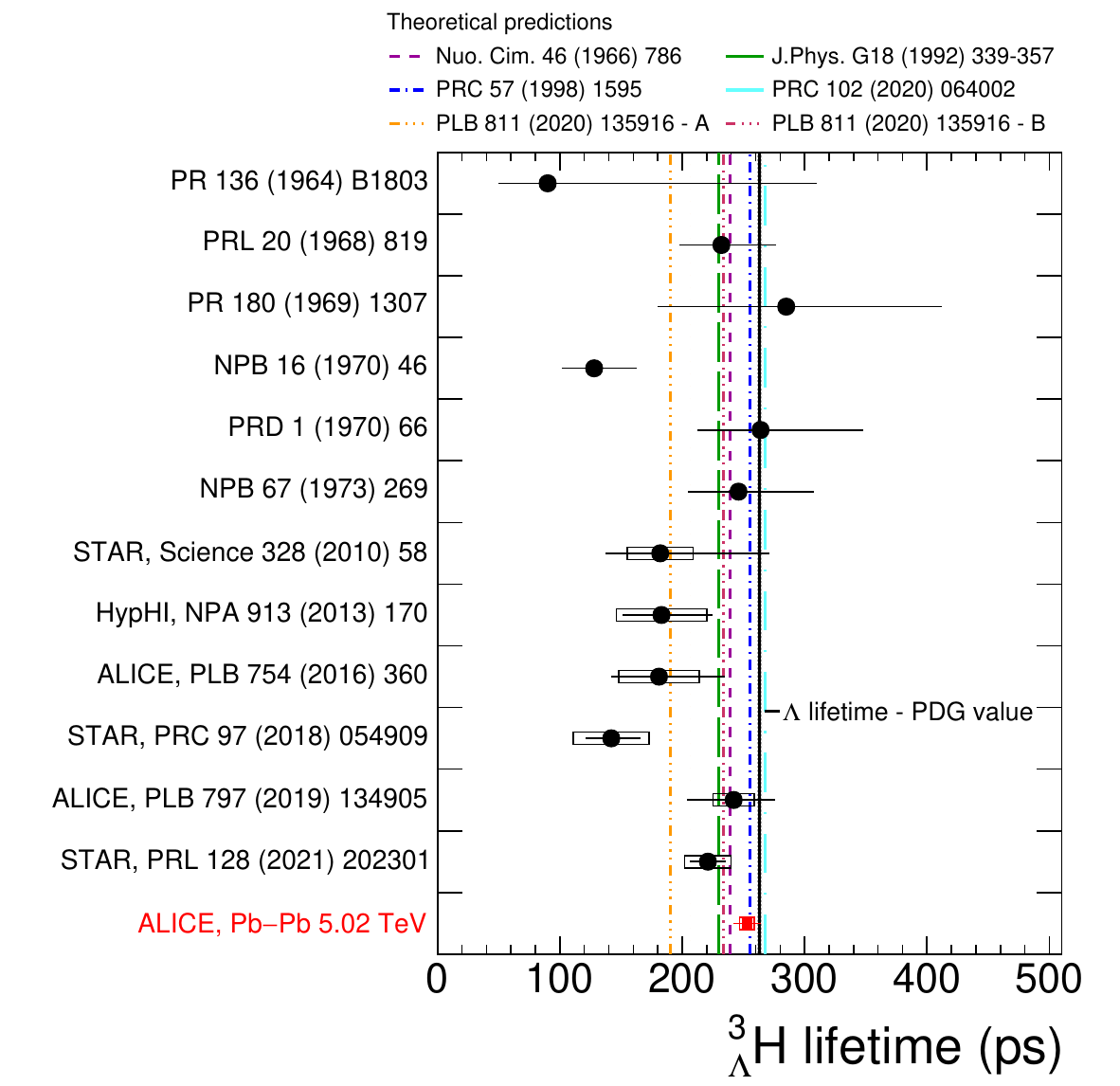}
    \includegraphics[width=0.49\textwidth]{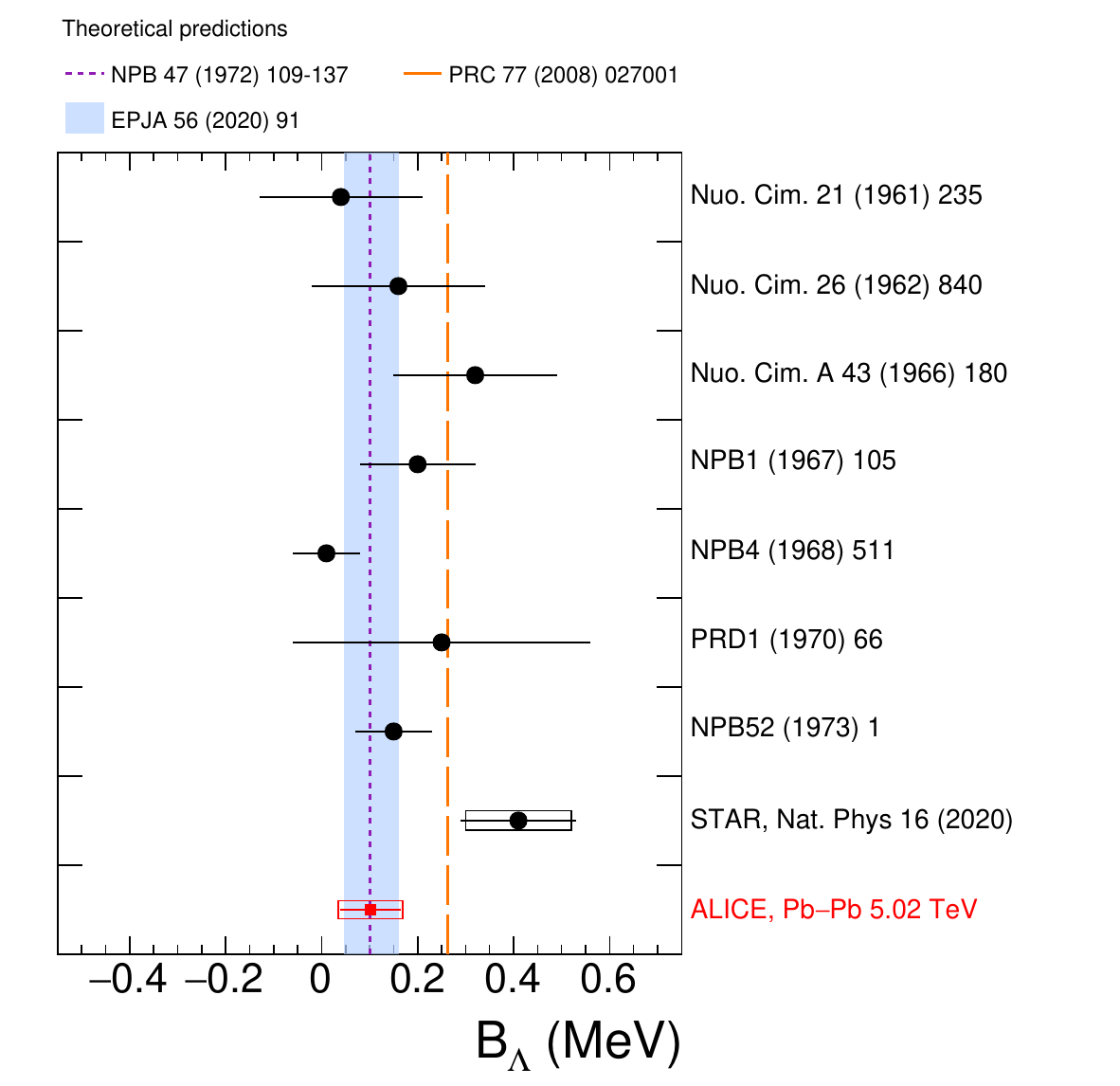}
    \caption{Collection of the \hyp lifetime (left)~\cite{PhysRev.136.B1803,Keyes:1968zz,Phillips:1969uy,Bohm:1970se,Keyes:1970ck,Keyes:1974ev,STAR:2010gyg,Rappold:2013fic,Adam:2015yta,Adamczyk:2017buv,Acharya:2019qcp,STAR:2021orx} and \BL (right)~\cite{Prakash1961OnTB,ammar1962,Mayeur1966ADO,Gajewski:1967ruj,Bohm:1968qkc,Keyes:1970ck,Juric:1973zq,Adam:2019phl} measurements obtained with different experimental techniques. The horizontal lines and boxes are the statistical and systematic uncertainties, respectively. The dashed and dash-dotted lines are the corresponding theoretical predictions~\cite{Congleton:1992kk,Kamada:1997rv,Rayet:1966fe,Hildenbrand:2020kzu,Perez-Obiol:2020qjy,Dalitz:1972vzj,Fujiwara:2007en,Haidenbauer:2019boi}.
    Two predictions are reported in~\cite{Perez-Obiol:2020qjy}: prediction A assumes \BL = 130 keV, while prediction B assumes \BL = 69 keV.}
    \label{fig:collections}
\end{figure*}

Even though some local tensions among a few measurements of lifetime and \BL have been reported in the literature as the "hypertriton puzzle", when performing a global average of the historically available measurements (see Fig.~\ref{fig:collections} and the Mainz hypernuclear data database~\cite{HypernuclearDataBase}), the probability of having such a set of measurements, computed with a Pearson test, is 23\% for the lifetime and 57\% for the \BL, hence no global tension is found.
A remaining piece to be set for the complete understanding of the \hyp structure is the measurement of branching ratios for the various decay channels~\cite{Hildenbrand:2020kzu}. The Run 3 of the LHC will make those measurements accessible with unprecedented precision.

%% file: fa_2022-08-31_Opt_C.tex

The ALICE Collaboration would like to thank all its engineers and technicians for their invaluable contributions to the construction of the experiment and the CERN accelerator teams for the outstanding performance of the LHC complex.
The ALICE Collaboration gratefully acknowledges the resources and support provided by all Grid centres and the Worldwide LHC Computing Grid (WLCG) collaboration.
The ALICE Collaboration acknowledges the following funding agencies for their support in building and running the ALICE detector:
A. I. Alikhanyan National Science Laboratory (Yerevan Physics Institute) Foundation (ANSL), State Committee of Science and World Federation of Scientists (WFS), Armenia;
Austrian Academy of Sciences, Austrian Science Fund (FWF): [M 2467-N36] and Nationalstiftung f\"{u}r Forschung, Technologie und Entwicklung, Austria;
Ministry of Communications and High Technologies, National Nuclear Research Center, Azerbaijan;
Conselho Nacional de Desenvolvimento Cient\'{\i}fico e Tecnol\'{o}gico (CNPq), Financiadora de Estudos e Projetos (Finep), Funda\c{c}\~{a}o de Amparo \`{a} Pesquisa do Estado de S\~{a}o Paulo (FAPESP) and Universidade Federal do Rio Grande do Sul (UFRGS), Brazil;
Bulgarian Ministry of Education and Science, within the National Roadmap for Research Infrastructures 2020-2027 (object CERN), Bulgaria;
Ministry of Education of China (MOEC) , Ministry of Science \& Technology of China (MSTC) and National Natural Science Foundation of China (NSFC), China;
Ministry of Science and Education and Croatian Science Foundation, Croatia;
Centro de Aplicaciones Tecnol\'{o}gicas y Desarrollo Nuclear (CEADEN), Cubaenerg\'{\i}a, Cuba;
Ministry of Education, Youth and Sports of the Czech Republic, Czech Republic;
The Danish Council for Independent Research | Natural Sciences, the VILLUM FONDEN and Danish National Research Foundation (DNRF), Denmark;
Helsinki Institute of Physics (HIP), Finland;
Commissariat \`{a} l'Energie Atomique (CEA) and Institut National de Physique Nucl\'{e}aire et de Physique des Particules (IN2P3) and Centre National de la Recherche Scientifique (CNRS), France;
Bundesministerium f\"{u}r Bildung und Forschung (BMBF) and GSI Helmholtzzentrum f\"{u}r Schwerionenforschung GmbH, Germany;
General Secretariat for Research and Technology, Ministry of Education, Research and Religions, Greece;
National Research, Development and Innovation Office, Hungary;
Department of Atomic Energy Government of India (DAE), Department of Science and Technology, Government of India (DST), University Grants Commission, Government of India (UGC) and Council of Scientific and Industrial Research (CSIR), India;
National Research and Innovation Agency - BRIN, Indonesia;
Istituto Nazionale di Fisica Nucleare (INFN), Italy;
Japanese Ministry of Education, Culture, Sports, Science and Technology (MEXT) and Japan Society for the Promotion of Science (JSPS) KAKENHI, Japan;
Consejo Nacional de Ciencia (CONACYT) y Tecnolog\'{i}a, through Fondo de Cooperaci\'{o}n Internacional en Ciencia y Tecnolog\'{i}a (FONCICYT) and Direcci\'{o}n General de Asuntos del Personal Academico (DGAPA), Mexico;
Nederlandse Organisatie voor Wetenschappelijk Onderzoek (NWO), Netherlands;
The Research Council of Norway, Norway;
Commission on Science and Technology for Sustainable Development in the South (COMSATS), Pakistan;
Pontificia Universidad Cat\'{o}lica del Per\'{u}, Peru;
Ministry of Education and Science, National Science Centre and WUT ID-UB, Poland;
Korea Institute of Science and Technology Information and National Research Foundation of Korea (NRF), Republic of Korea;
Ministry of Education and Scientific Research, Institute of Atomic Physics, Ministry of Research and Innovation and Institute of Atomic Physics and University Politehnica of Bucharest, Romania;
Ministry of Education, Science, Research and Sport of the Slovak Republic, Slovakia;
National Research Foundation of South Africa, South Africa;
Swedish Research Council (VR) and Knut \& Alice Wallenberg Foundation (KAW), Sweden;
European Organization for Nuclear Research, Switzerland;
Suranaree University of Technology (SUT), National Science and Technology Development Agency (NSTDA), Thailand Science Research and Innovation (TSRI) and National Science, Research and Innovation Fund (NSRF), Thailand;
Turkish Energy, Nuclear and Mineral Research Agency (TENMAK), Turkey;
National Academy of  Sciences of Ukraine, Ukraine;
Science and Technology Facilities Council (STFC), United Kingdom;
National Science Foundation of the United States of America (NSF) and United States Department of Energy, Office of Nuclear Physics (DOE NP), United States of America.
In addition, individual groups or members have received support from:
Marie Sk\l{}odowska Curie, European Research Council, Strong 2020 - Horizon 2020 (grant nos. 950692, 824093, 896850), European Union;
Academy of Finland (Center of Excellence in Quark Matter) (grant nos. 346327, 346328), Finland;
Programa de Apoyos para la Superaci\'{o}n del Personal Acad\'{e}mico, UNAM, Mexico.

%% file: appendix.tex
\section{Additional Figures}
\label{suppmat}
\begin{figure*}[h!]
    \centering
    \includegraphics[width=0.3\textwidth]{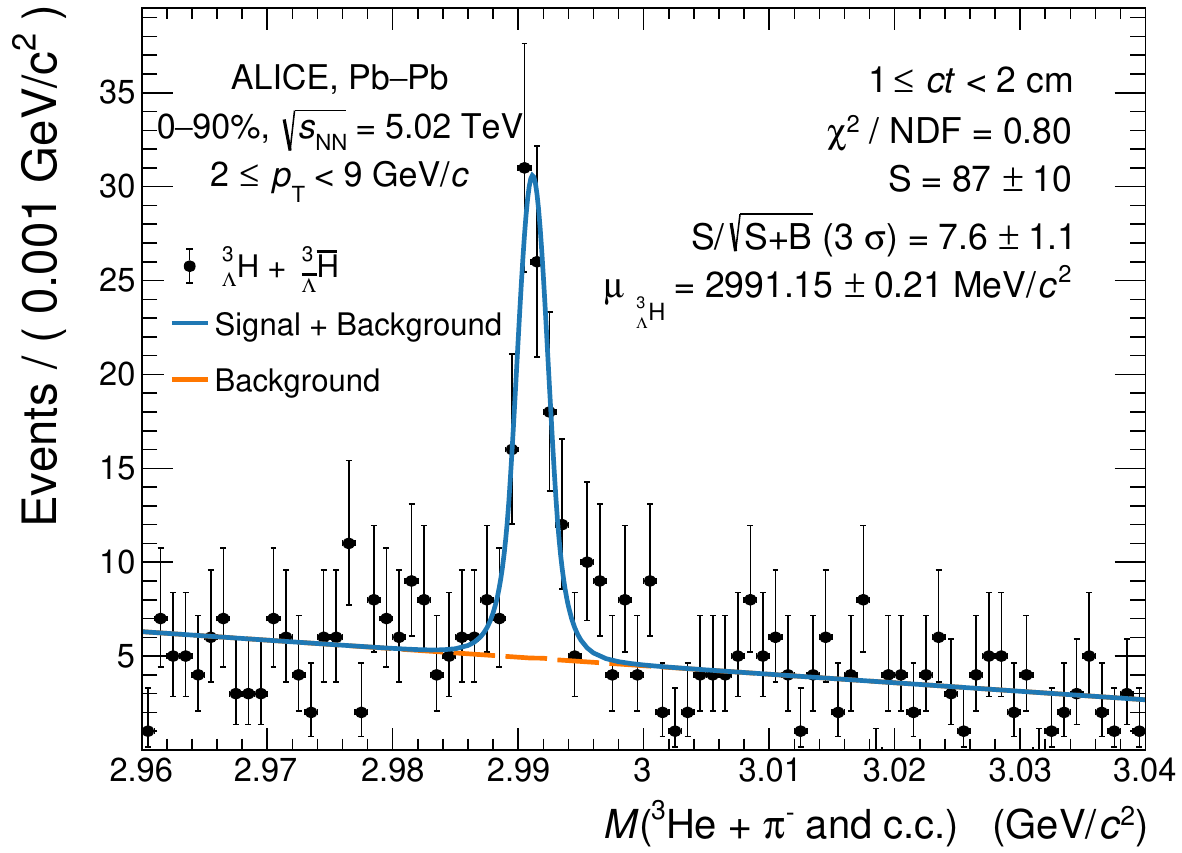}
    \includegraphics[width=0.3\textwidth]{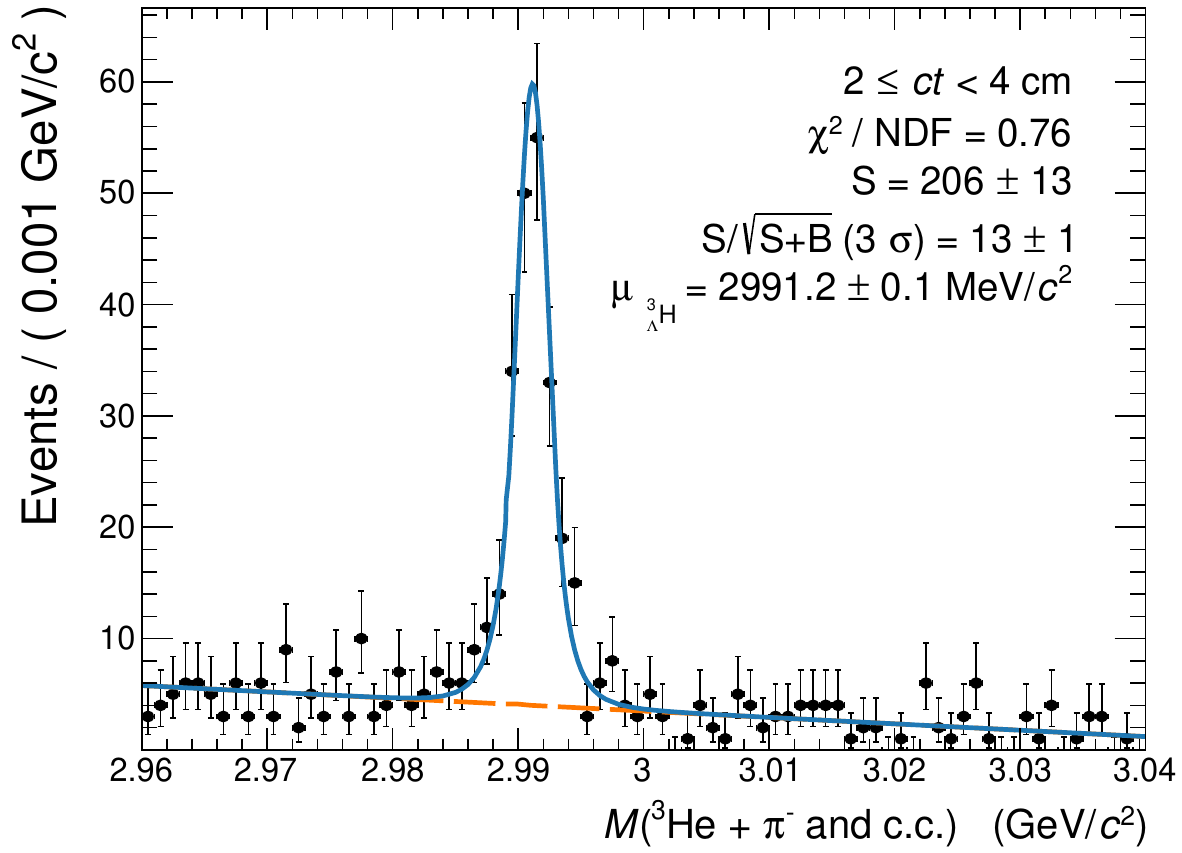}
    \includegraphics[width=0.3\textwidth]{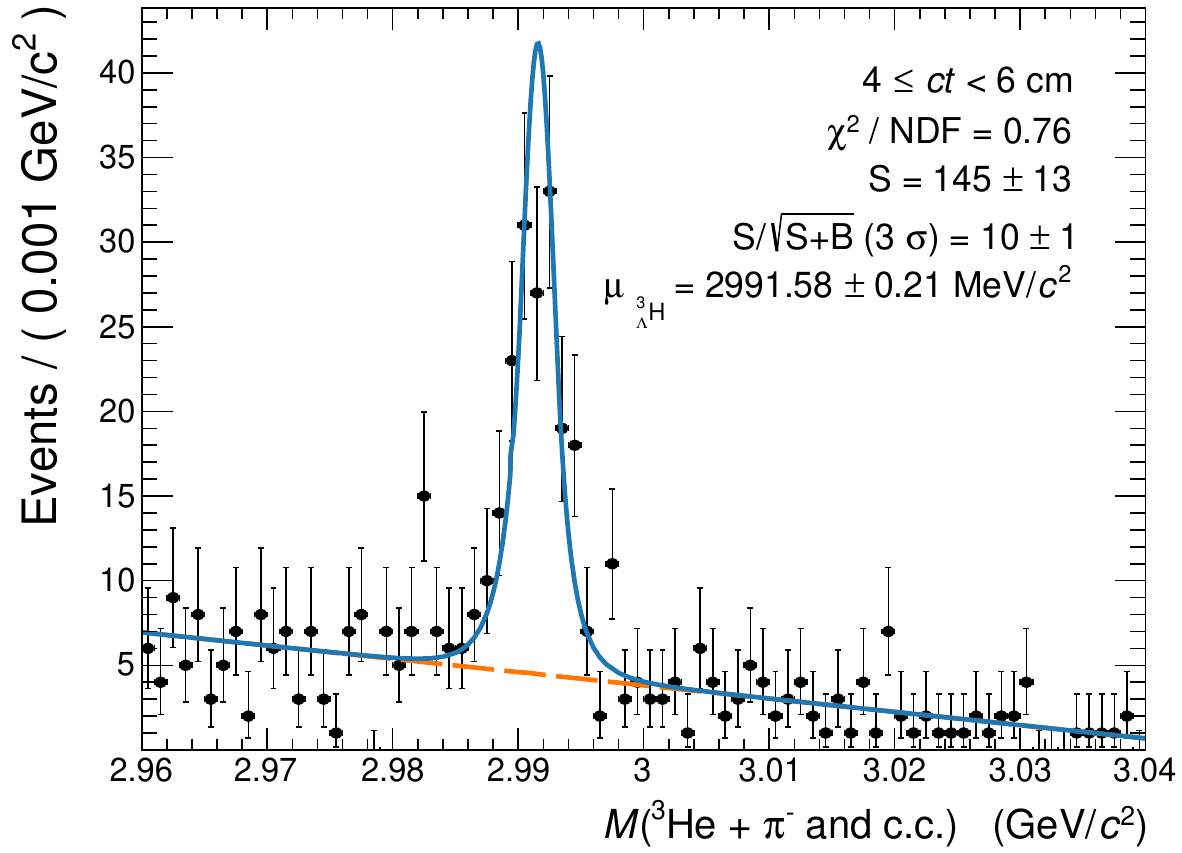}
    \includegraphics[width=0.3\textwidth]{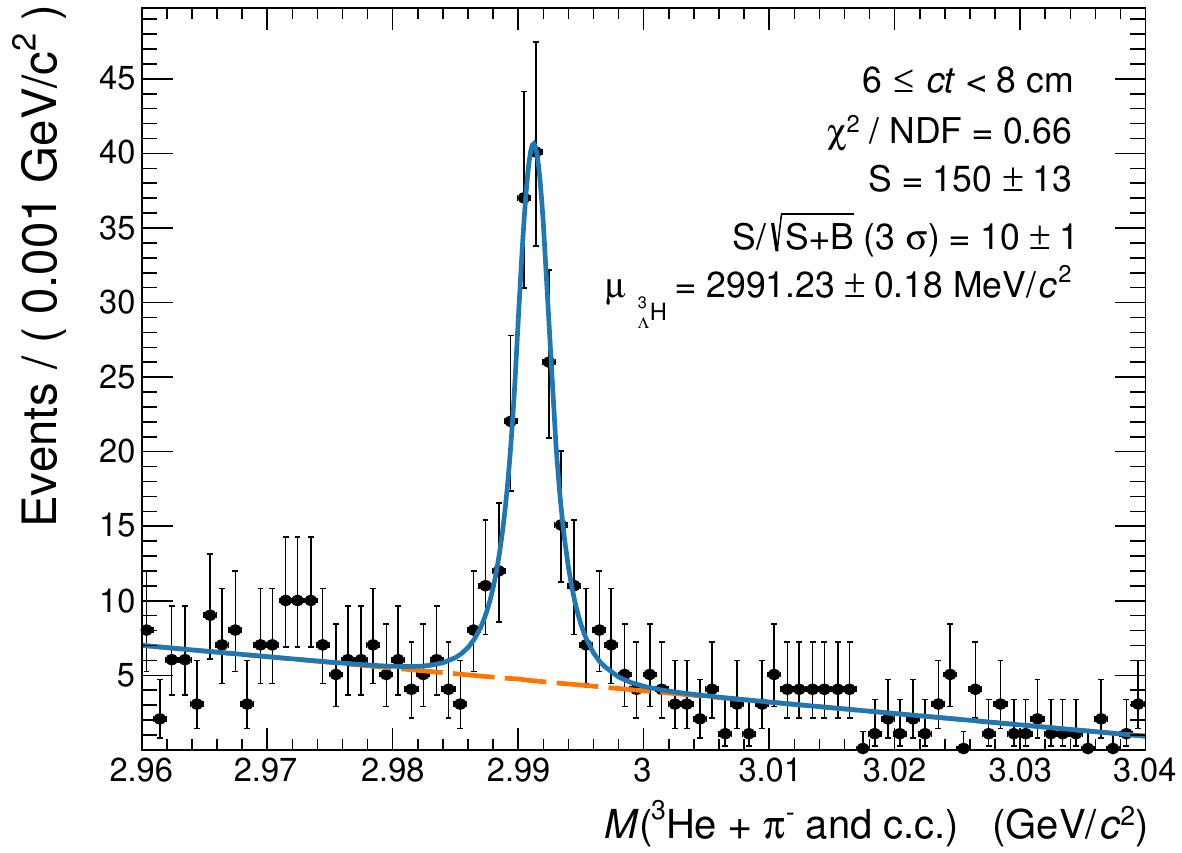}
    \includegraphics[width=0.3\textwidth]{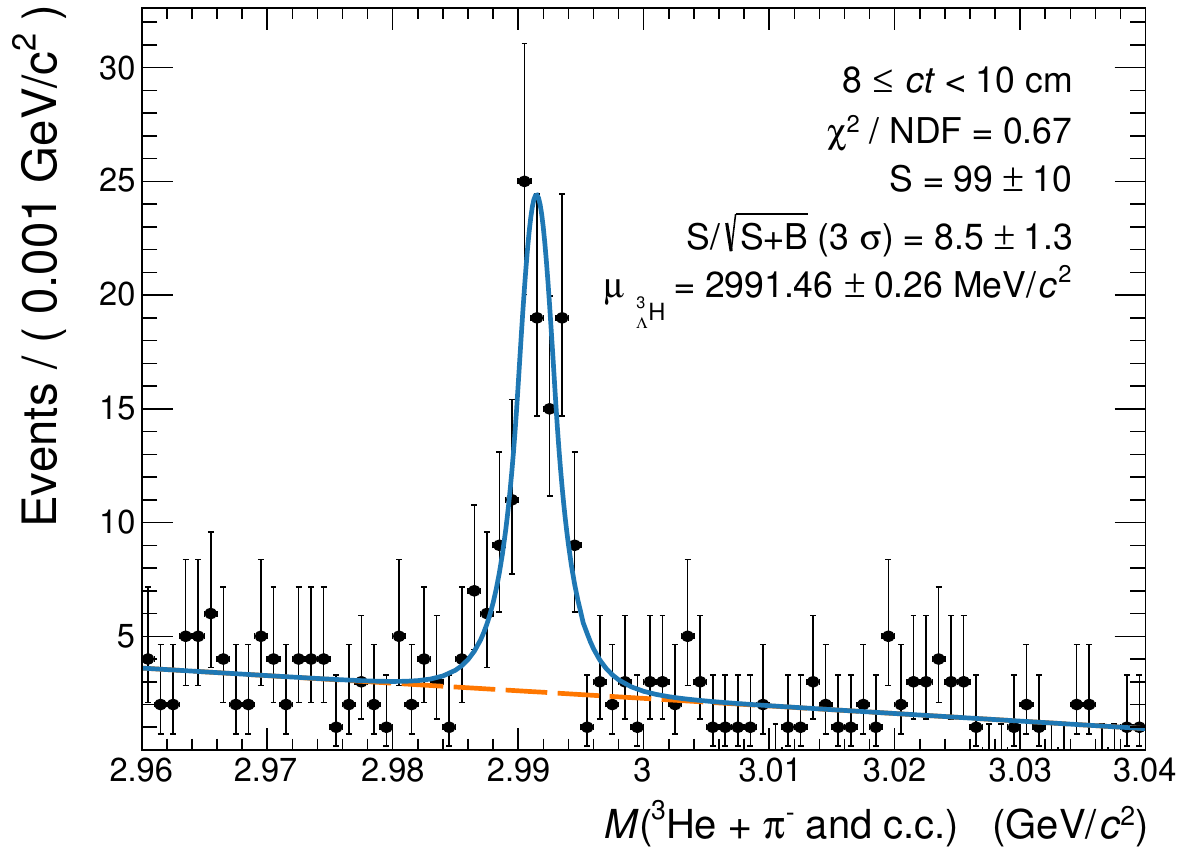}
    \includegraphics[width=0.3\textwidth]{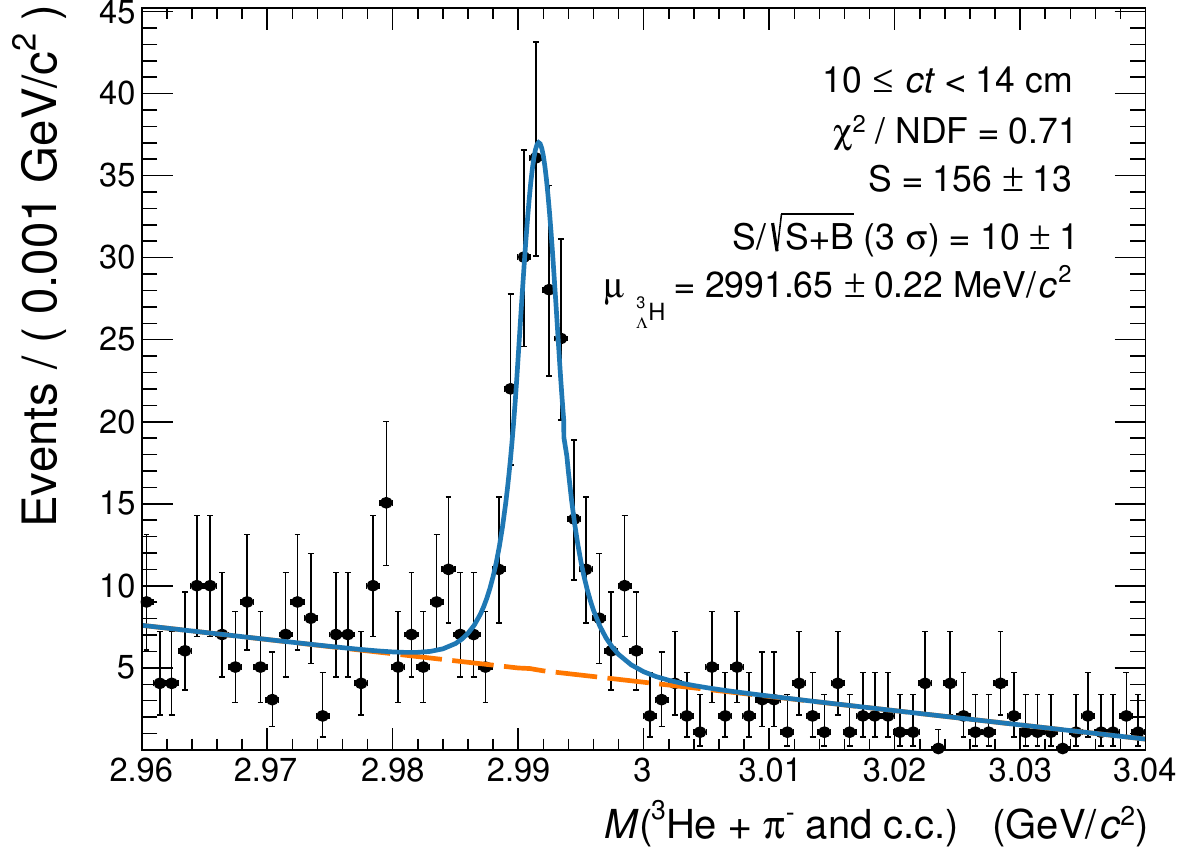}
    \includegraphics[width=0.3\textwidth]{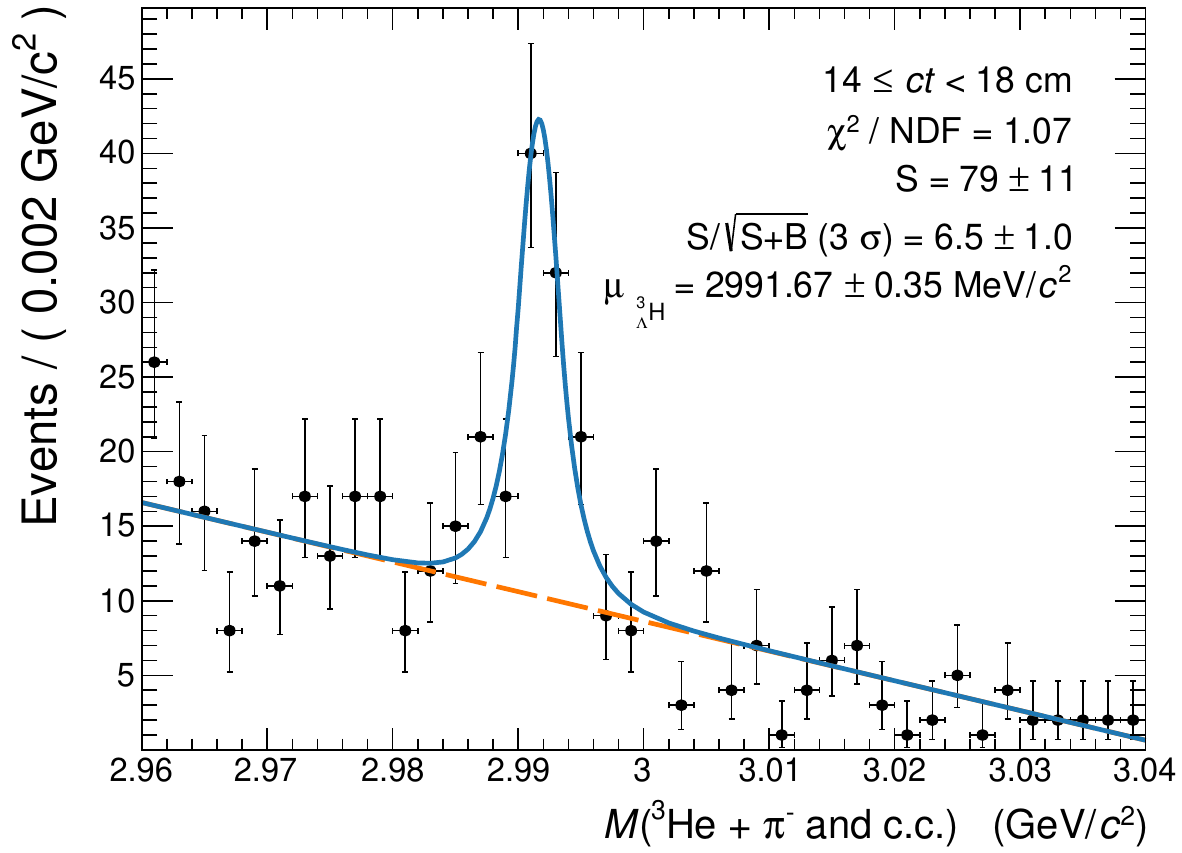}
    \includegraphics[width=0.3\textwidth]{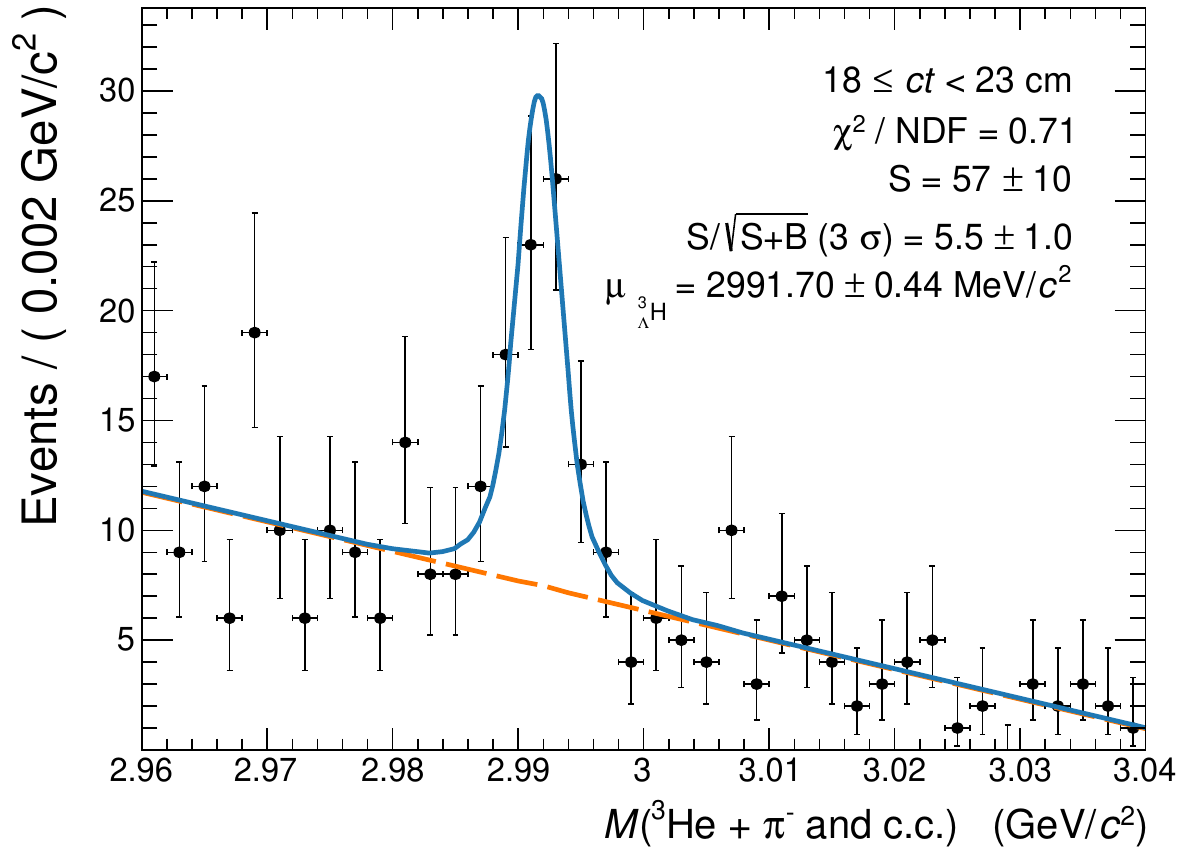}
    \includegraphics[width=0.3\textwidth]{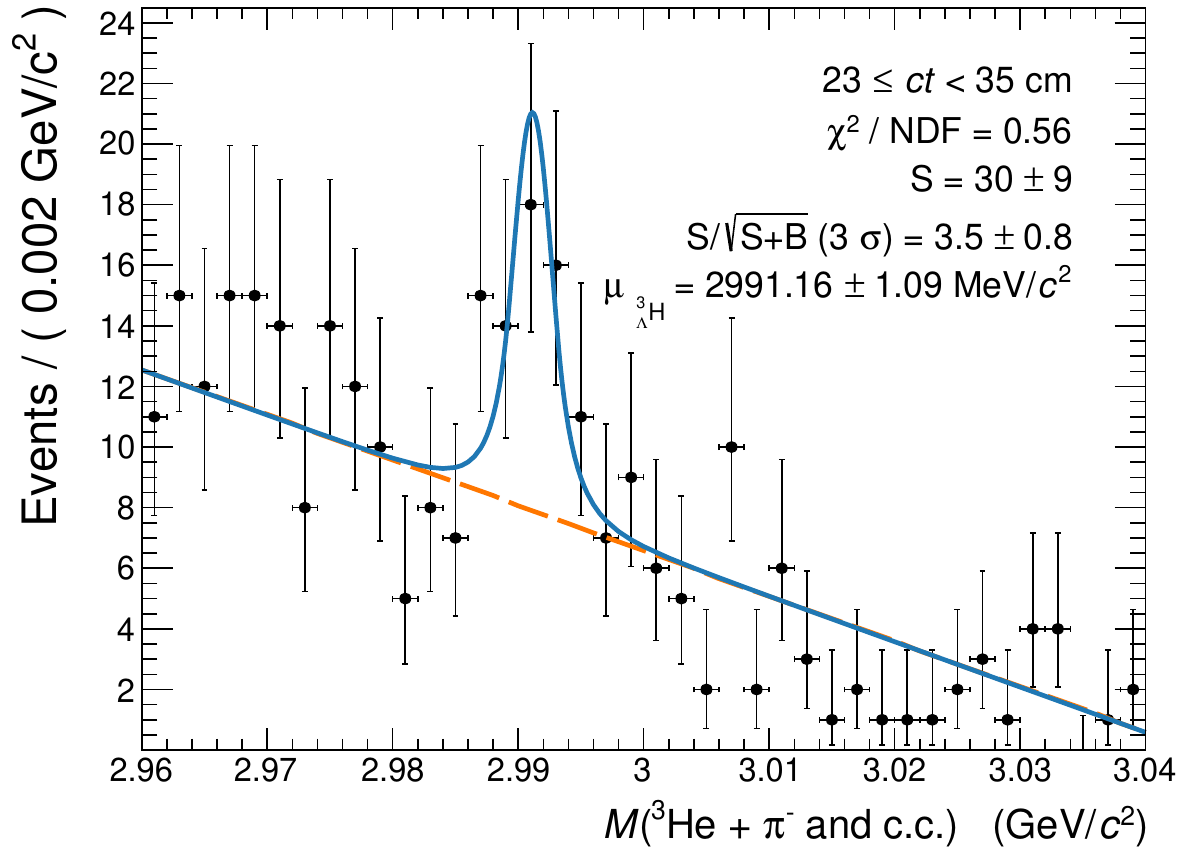}
    \caption{Distribution of the invariant mass of the \hyp and \antihyp candidates in nine \ct intervals from 1 to 35 cm. The statistical uncertainties of the bin counts are represented with vertical lines. The distribution is fitted with a two-component model; the blue line depicts the overall fit, and the orange dashed line displays the background component.}
    \label{fig:invmass}
\end{figure*}

\begin{figure*}[h!]
    \centering
    \includegraphics[width=0.5\textwidth]{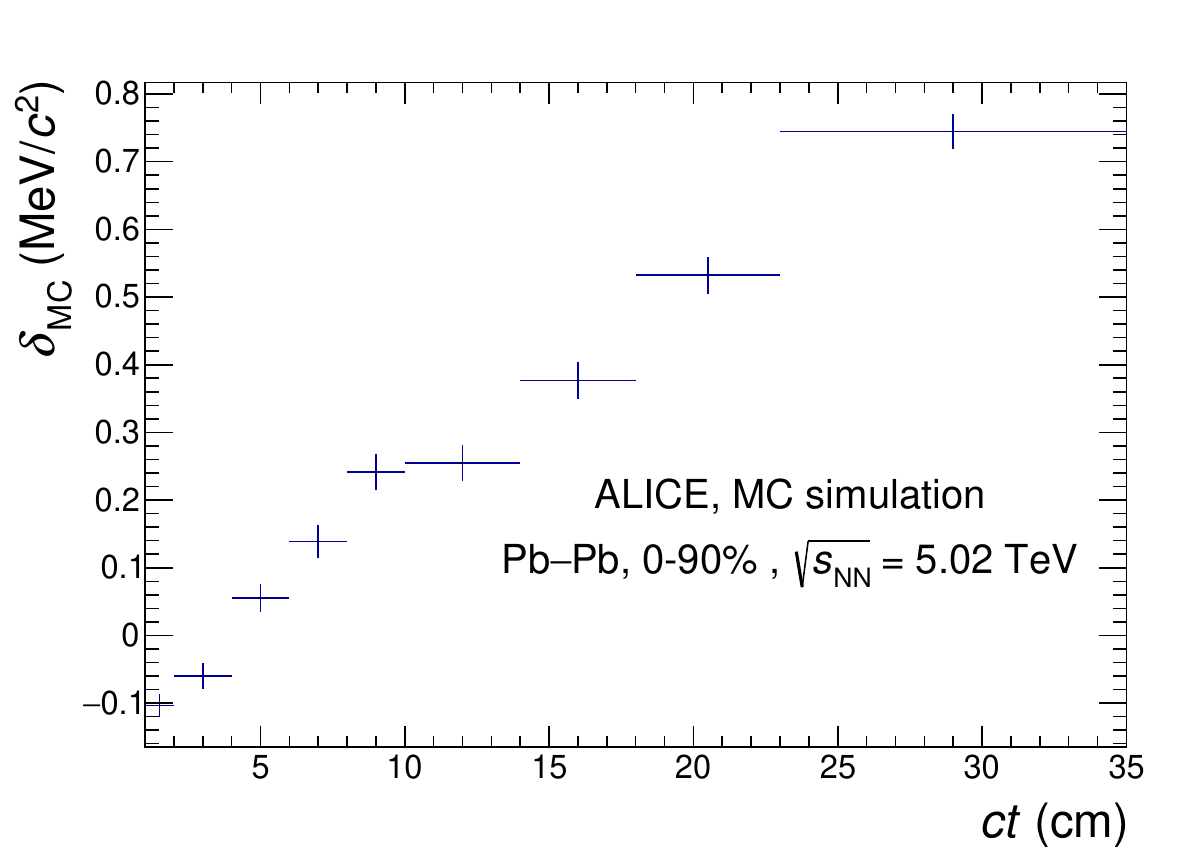}
    \caption{Reconstruction shift $\delta_\mathrm{MC}$ observed in the MC as a function of the proper decay length. The statistical uncertainties are represented with vertical lines. The value of $\delta_\mathrm{MC}$ increases with the distance travelled by the \hyp candidates before decaying.}
    \label{fig:recoshift}
\end{figure*}

%% file: 2022-08-31-Alice_Authorlist_2022-08-31_Opt_C.tex
\begin{flushleft} 
\small

S.~Acharya\,\orcidlink{0000-0002-9213-5329}\,$^{\rm 124}$, 
D.~Adamov\'{a}\,\orcidlink{0000-0002-0504-7428}\,$^{\rm 86}$, 
A.~Adler$^{\rm 69}$, 
G.~Aglieri Rinella\,\orcidlink{0000-0002-9611-3696}\,$^{\rm 32}$, 
M.~Agnello\,\orcidlink{0000-0002-0760-5075}\,$^{\rm 29}$, 
N.~Agrawal\,\orcidlink{0000-0003-0348-9836}\,$^{\rm 50}$, 
Z.~Ahammed\,\orcidlink{0000-0001-5241-7412}\,$^{\rm 131}$, 
S.~Ahmad\,\orcidlink{0000-0003-0497-5705}\,$^{\rm 15}$, 
S.U.~Ahn\,\orcidlink{0000-0001-8847-489X}\,$^{\rm 70}$, 
I.~Ahuja\,\orcidlink{0000-0002-4417-1392}\,$^{\rm 37}$
% ...

\section*{Affiliation Notes}

$^{\rm I}$ Deceased\\
$^{\rm II}$ Also at: Max-Planck-Institut f\"{u}r Physik, Munich, Germany\\
$^{\rm III}$ Also at: Italian National Agency for New Technologies, Energy and Sustainable Economic Development (ENEA), Bologna, Italy\\
$^{\rm IV}$ Also at: Dipartimento DET del Politecnico di Torino, Turin, Italy\\
$^{\rm V}$ Also at: Department of Applied Physics, Aligarh Muslim University, Aligarh, India\\
$^{\rm VI}$ Also at: Institute of Theoretical Physics, University of Wroclaw, Poland\\
$^{\rm VII}$ Also at: An institution covered by a cooperation agreement with CERN\\

\section*{Collaboration Institutes}

$^{1}$ A.I. Alikhanyan National Science Laboratory (Yerevan Physics Institute) Foundation, Yerevan, Armenia\\
$^{2}$ AGH University of Science and Technology, Cracow, Poland\\
$^{3}$ Bogolyubov Institute for Theoretical Physics, National Academy of Sciences of Ukraine, Kiev, Ukraine\\
$^{4}$ Bose Institute, Department of Physics  and Centre for Astroparticle Physics and Space Science (CAPSS), Kolkata, India\\
$^{5}$ California Polytechnic State University, San Luis Obispo, California, United States\\
$^{6}$ Central China Normal University, Wuhan, China\\
$^{7}$ Centro de Aplicaciones Tecnol\'{o}gicas y Desarrollo Nuclear (CEADEN), Havana, Cuba\\
$^{8}$ Centro de Investigaci\'{o}n y de Estudios Avanzados (CINVESTAV), Mexico City and M\'{e}rida, Mexico\\
$^{9}$ Chicago State University, Chicago, Illinois, United States\\
$^{10}$ China Institute of Atomic Energy, Beijing, China\\
$^{11}$ Chungbuk National University, Cheongju, Republic of Korea\\
$^{12}$ Comenius University Bratislava, Faculty of Mathematics, Physics and Informatics, Bratislava, Slovak Republic\\
$^{13}$ COMSATS University Islamabad, Islamabad, Pakistan\\
$^{14}$ Creighton University, Omaha, Nebraska, United States\\
$^{15}$ Department of Physics, Aligarh Muslim University, Aligarh, India\\
$^{16}$ Department of Physics, Pusan National University, Pusan, Republic of Korea\\
$^{17}$ Department of Physics, Sejong University, Seoul, Republic of Korea\\
$^{18}$ Department of Physics, University of California, Berkeley, California, United States\\
$^{19}$ Department of Physics, University of Oslo, Oslo, Norway\\
$^{20}$ Department of Physics and Technology, University of Bergen, Bergen, Norway\\
$^{21}$ Dipartimento di Fisica, Universit\`{a} di Pavia, Pavia, Italy\\
$^{22}$ Dipartimento di Fisica dell'Universit\`{a} and Sezione INFN, Cagliari, Italy\\
$^{23}$ Dipartimento di Fisica dell'Universit\`{a} and Sezione INFN, Trieste, Italy\\
$^{24}$ Dipartimento di Fisica dell'Universit\`{a} and Sezione INFN, Turin, Italy\\
$^{25}$ Dipartimento di Fisica e Astronomia dell'Universit\`{a} and Sezione INFN, Bologna, Italy\\
$^{26}$ Dipartimento di Fisica e Astronomia dell'Universit\`{a} and Sezione INFN, Catania, Italy\\
$^{27}$ Dipartimento di Fisica e Astronomia dell'Universit\`{a} and Sezione INFN, Padova, Italy\\
$^{28}$ Dipartimento di Fisica `E.R.~Caianiello' dell'Universit\`{a} and Gruppo Collegato INFN, Salerno, Italy\\
$^{29}$ Dipartimento DISAT del Politecnico and Sezione INFN, Turin, Italy\\
$^{30}$ Dipartimento di Scienze MIFT, Universit\`{a} di Messina, Messina, Italy\\
$^{31}$ Dipartimento Interateneo di Fisica `M.~Merlin' and Sezione INFN, Bari, Italy\\
$^{32}$ European Organization for Nuclear Research (CERN), Geneva, Switzerland\\
$^{33}$ Faculty of Electrical Engineering, Mechanical Engineering and Naval Architecture, University of Split, Split, Croatia\\
$^{34}$ Faculty of Engineering and Science, Western Norway University of Applied Sciences, Bergen, Norway\\
$^{35}$ Faculty of Nuclear Sciences and Physical Engineering, Czech Technical University in Prague, Prague, Czech Republic\\
$^{36}$ Faculty of Physics, Sofia University, Sofia, Bulgaria\\
$^{37}$ Faculty of Science, P.J.~\v{S}af\'{a}rik University, Ko\v{s}ice, Slovak Republic\\
$^{38}$ Frankfurt Institute for Advanced Studies, Johann Wolfgang Goethe-Universit\"{a}t Frankfurt, Frankfurt, Germany\\
$^{39}$ Fudan University, Shanghai, China\\
$^{40}$ Gangneung-Wonju National University, Gangneung, Republic of Korea\\
$^{41}$ Gauhati University, Department of Physics, Guwahati, India\\
$^{42}$ Helmholtz-Institut f\"{u}r Strahlen- und Kernphysik, Rheinische Friedrich-Wilhelms-Universit\"{a}t Bonn, Bonn, Germany\\
$^{43}$ Helsinki Institute of Physics (HIP), Helsinki, Finland\\
$^{44}$ High Energy Physics Group,  Universidad Aut\'{o}noma de Puebla, Puebla, Mexico\\
$^{45}$ Horia Hulubei National Institute of Physics and Nuclear Engineering, Bucharest, Romania\\
$^{46}$ Indian Institute of Technology Bombay (IIT), Mumbai, India\\
$^{47}$ Indian Institute of Technology Indore, Indore, India\\
$^{48}$ INFN, Laboratori Nazionali di Frascati, Frascati, Italy\\
$^{49}$ INFN, Sezione di Bari, Bari, Italy\\
$^{50}$ INFN, Sezione di Bologna, Bologna, Italy\\
$^{51}$ INFN, Sezione di Cagliari, Cagliari, Italy\\
$^{52}$ INFN, Sezione di Catania, Catania, Italy\\
$^{53}$ INFN, Sezione di Padova, Padova, Italy\\
$^{54}$ INFN, Sezione di Pavia, Pavia, Italy\\
$^{55}$ INFN, Sezione di Torino, Turin, Italy\\
$^{56}$ INFN, Sezione di Trieste, Trieste, Italy\\
$^{57}$ Inha University, Incheon, Republic of Korea\\
$^{58}$ Institute for Gravitational and Subatomic Physics (GRASP), Utrecht University/Nikhef, Utrecht, Netherlands\\
$^{59}$ Institute of Experimental Physics, Slovak Academy of Sciences, Ko\v{s}ice, Slovak Republic\\
$^{60}$ Institute of Physics, Homi Bhabha National Institute, Bhubaneswar, India\\
$^{61}$ Institute of Physics of the Czech Academy of Sciences, Prague, Czech Republic\\
$^{62}$ Institute of Space Science (ISS), Bucharest, Romania\\
$^{63}$ Institut f\"{u}r Kernphysik, Johann Wolfgang Goethe-Universit\"{a}t Frankfurt, Frankfurt, Germany\\
$^{64}$ Instituto de Ciencias Nucleares, Universidad Nacional Aut\'{o}noma de M\'{e}xico, Mexico City, Mexico\\
$^{65}$ Instituto de F\'{i}sica, Universidade Federal do Rio Grande do Sul (UFRGS), Porto Alegre, Brazil\\
$^{66}$ Instituto de F\'{\i}sica, Universidad Nacional Aut\'{o}noma de M\'{e}xico, Mexico City, Mexico\\
$^{67}$ iThemba LABS, National Research Foundation, Somerset West, South Africa\\
$^{68}$ Jeonbuk National University, Jeonju, Republic of Korea\\
$^{69}$ Johann-Wolfgang-Goethe Universit\"{a}t Frankfurt Institut f\"{u}r Informatik, Fachbereich Informatik und Mathematik, Frankfurt, Germany\\
$^{70}$ Korea Institute of Science and Technology Information, Daejeon, Republic of Korea\\
$^{71}$ KTO Karatay University, Konya, Turkey\\
$^{72}$ Laboratoire de Physique des 2 Infinis, Ir\`{e}ne Joliot-Curie, Orsay, France\\
$^{73}$ Laboratoire de Physique Subatomique et de Cosmologie, Universit\'{e} Grenoble-Alpes, CNRS-IN2P3, Grenoble, France\\
$^{74}$ Lawrence Berkeley National Laboratory, Berkeley, California, United States\\
$^{75}$ Lund University Department of Physics, Division of Particle Physics, Lund, Sweden\\
$^{76}$ Nagasaki Institute of Applied Science, Nagasaki, Japan\\
$^{77}$ Nara Women{'}s University (NWU), Nara, Japan\\
$^{78}$ National and Kapodistrian University of Athens, School of Science, Department of Physics , Athens, Greece\\
$^{79}$ National Centre for Nuclear Research, Warsaw, Poland\\
$^{80}$ National Institute of Science Education and Research, Homi Bhabha National Institute, Jatni, India\\
$^{81}$ National Nuclear Research Center, Baku, Azerbaijan\\
$^{82}$ National Research and Innovation Agency - BRIN, Jakarta, Indonesia\\
$^{83}$ Niels Bohr Institute, University of Copenhagen, Copenhagen, Denmark\\
$^{84}$ Nikhef, National institute for subatomic physics, Amsterdam, Netherlands\\
$^{85}$ Nuclear Physics Group, STFC Daresbury Laboratory, Daresbury, United Kingdom\\
$^{86}$ Nuclear Physics Institute of the Czech Academy of Sciences, Husinec-\v{R}e\v{z}, Czech Republic\\
$^{87}$ Oak Ridge National Laboratory, Oak Ridge, Tennessee, United States\\
$^{88}$ Ohio State University, Columbus, Ohio, United States\\
$^{89}$ Physics department, Faculty of science, University of Zagreb, Zagreb, Croatia\\
$^{90}$ Physics Department, Panjab University, Chandigarh, India\\
$^{91}$ Physics Department, University of Jammu, Jammu, India\\
$^{92}$ Physics Program and International Institute for Sustainability with Knotted Chiral Meta Matter (SKCM2), Hiroshima University, Hiroshima, Japan\\
$^{93}$ Physikalisches Institut, Eberhard-Karls-Universit\"{a}t T\"{u}bingen, T\"{u}bingen, Germany\\
$^{94}$ Physikalisches Institut, Ruprecht-Karls-Universit\"{a}t Heidelberg, Heidelberg, Germany\\
$^{95}$ Physik Department, Technische Universit\"{a}t M\"{u}nchen, Munich, Germany\\
$^{96}$ Politecnico di Bari and Sezione INFN, Bari, Italy\\
$^{97}$ Research Division and ExtreMe Matter Institute EMMI, GSI Helmholtzzentrum f\"ur Schwerionenforschung GmbH, Darmstadt, Germany\\
$^{98}$ Saga University, Saga, Japan\\
$^{99}$ Saha Institute of Nuclear Physics, Homi Bhabha National Institute, Kolkata, India\\
$^{100}$ School of Physics and Astronomy, University of Birmingham, Birmingham, United Kingdom\\
$^{101}$ Secci\'{o}n F\'{\i}sica, Departamento de Ciencias, Pontificia Universidad Cat\'{o}lica del Per\'{u}, Lima, Peru\\
$^{102}$ Stefan Meyer Institut f\"{u}r Subatomare Physik (SMI), Vienna, Austria\\
$^{103}$ SUBATECH, IMT Atlantique, Nantes Universit\'{e}, CNRS-IN2P3, Nantes, France\\
$^{104}$ Suranaree University of Technology, Nakhon Ratchasima, Thailand\\
$^{105}$ Technical University of Ko\v{s}ice, Ko\v{s}ice, Slovak Republic\\
$^{106}$ The Henryk Niewodniczanski Institute of Nuclear Physics, Polish Academy of Sciences, Cracow, Poland\\
$^{107}$ The University of Texas at Austin, Austin, Texas, United States\\
$^{108}$ Universidad Aut\'{o}noma de Sinaloa, Culiac\'{a}n, Mexico\\
$^{109}$ Universidade de S\~{a}o Paulo (USP), S\~{a}o Paulo, Brazil\\
$^{110}$ Universidade Estadual de Campinas (UNICAMP), Campinas, Brazil\\
$^{111}$ Universidade Federal do ABC, Santo Andre, Brazil\\
$^{112}$ University of Cape Town, Cape Town, South Africa\\
$^{113}$ University of Houston, Houston, Texas, United States\\
$^{114}$ University of Jyv\"{a}skyl\"{a}, Jyv\"{a}skyl\"{a}, Finland\\
$^{115}$ University of Kansas, Lawrence, Kansas, United States\\
$^{116}$ University of Liverpool, Liverpool, United Kingdom\\
$^{117}$ University of Science and Technology of China, Hefei, China\\
$^{118}$ University of South-Eastern Norway, Kongsberg, Norway\\
$^{119}$ University of Tennessee, Knoxville, Tennessee, United States\\
$^{120}$ University of the Witwatersrand, Johannesburg, South Africa\\
$^{121}$ University of Tokyo, Tokyo, Japan\\
$^{122}$ University of Tsukuba, Tsukuba, Japan\\
$^{123}$ University Politehnica of Bucharest, Bucharest, Romania\\
$^{124}$ Universit\'{e} Clermont Auvergne, CNRS/IN2P3, LPC, Clermont-Ferrand, France\\
$^{125}$ Universit\'{e} de Lyon, CNRS/IN2P3, Institut de Physique des 2 Infinis de Lyon, Lyon, France\\
$^{126}$ Universit\'{e} de Strasbourg, CNRS, IPHC UMR 7178, F-67000 Strasbourg, France, Strasbourg, France\\
$^{127}$ Universit\'{e} Paris-Saclay Centre d'Etudes de Saclay (CEA), IRFU, D\'{e}partment de Physique Nucl\'{e}aire (DPhN), Saclay, France\\
$^{128}$ Universit\`{a} degli Studi di Foggia, Foggia, Italy\\
$^{129}$ Universit\`{a} del Piemonte Orientale, Vercelli, Italy\\
$^{130}$ Universit\`{a} di Brescia, Brescia, Italy\\
$^{131}$ Variable Energy Cyclotron Centre, Homi Bhabha National Institute, Kolkata, India\\
$^{132}$ Warsaw University of Technology, Warsaw, Poland\\
$^{133}$ Wayne State University, Detroit, Michigan, United States\\
$^{134}$ Westf\"{a}lische Wilhelms-Universit\"{a}t M\"{u}nster, Institut f\"{u}r Kernphysik, M\"{u}nster, Germany\\
$^{135}$ Wigner Research Centre for Physics, Budapest, Hungary\\
$^{136}$ Yale University, New Haven, Connecticut, United States\\
$^{137}$ Yonsei University, Seoul, Republic of Korea\\
$^{138}$  Zentrum  f\"{u}r Technologie und Transfer (ZTT), Worms, Germany\\
$^{139}$ Affiliated with an institute covered by a cooperation agreement with CERN\\
$^{140}$ Affiliated with an international laboratory covered by a cooperation agreement with CERN.\\

\end{flushleft} 